\documentclass[sigconf,authorversion]{acmart}

\AtBeginDocument{%
  \providecommand\BibTeX{{%
    \normalfont B\kern-0.5em{\scshape i\kern-0.25em b}\kern-0.8em\TeX}}}

\copyrightyear{2020} 
\acmYear{2020} 
\setcopyright{acmlicensed}\acmConference[DATA '20]{The 3rd International SenSys+BuildSys Workshop on Data: Acquisition to Analysis}{November 16--19, 2020}{Virtual Event, Japan}
\acmBooktitle{The 3rd International SenSys+BuildSys Workshop on Data: Acquisition to Analysis (DATA '20), November 16--19, 2020, Virtual Event, Japan}
\acmPrice{15.00}
\acmDOI{10.1145/3419016.3431491}
\acmISBN{978-1-4503-8136-9/20/11}

\usepackage{tabularx}
\usepackage{booktabs}
\usepackage{xspace}
\usepackage{paralist}

\newif\ifshowcomments
\showcommentstrue

\ifshowcomments
\newcommand{\mynote}[2]{\textcolor{blue}{\fbox{\bfseries\sffamily\scriptsize#1}}
  \textcolor{blue}{{$/*$\textsf{\emph{#2}}$*/$}}}
\else
\newcommand{\mynote}[2]{}
\fi

\newcommand{\loed}{\textsc{LoED}\xspace}

\newcommand{\fakeparagraph}[1]{\vspace{1mm}\noindent\textbf{#1.}}

\begin{document}

\title{Dataset: LoED: The LoRaWAN at the Edge Dataset}



\author{Laksh Bhatia}
\affiliation{%
  \institution{Imperial College London}
  \streetaddress{South Kensington Campus}
  \city{London}
  \state{UK}
  \postcode{SW7 2AZ}
}
\email{laksh.bhatia16@imperial.ac.uk}
\author{Michael Breza}
\affiliation{%
  \institution{Imperial College London}
  \streetaddress{South Kensington Campus}
  \city{London}
  \state{UK}
  \postcode{SW7 2AZ}
}
\email{michael.breza04@imperial.ac.uk}

\author{Ramona Marfievici}
\affiliation{%
  \institution{Digital Catapult}
    \city{London}
    \state{UK}
}
\email{ramona.marfievici@digicatapult.org.uk}
\author{Julie A. McCann}
\affiliation{%
\institution{Imperial College London}
  \streetaddress{South Kensington Campus}
  \city{London}
  \state{UK}
  \postcode{SW7 2AZ}
}
\email{j.mccann@imperial.ac.uk}
\renewcommand{\shortauthors}{L. Bhatia, et al.}

\begin{abstract}
    This paper presents the LoRaWAN at the Edge Dataset (LoED), an open LoRaWAN
    packet dataset collected at gateways.  Real-world LoRaWAN datasets
    are important for repeatable sensor-network and communications research and evaluation as, if carefully collected, they provide realistic
    working assumptions.  LoED data is collected from nine gateways over a four
    month period in a dense urban environment. The dataset contains
    packet header information and all physical layer properties reported by gateways such as the CRC, RSSI, SNR and spreading factor.
    Files are provided to analyse the data and get aggregated statistics. The dataset is available at: \url{doi.org/10.5281/zenodo.4121430}
\end{abstract}



\maketitle

\section{Introduction}

LoRaWAN is a wireless single-hop, long-range and Low-Power Wide Area Network (LPWAN). LoRaWAN has seen rapid adoption  due to its
communication coverage, ease of deployment, and simplified infrastructure management. Over $180$~million LoRa-enabled
devices\footnote{https://www.semtech.com/lora} are already used for a variety
of smart city and rural applications.

A significant body of academic work on LoRa and LoRaWAN has proposed improvements to data rate mechanism control~\cite{8271941} and solutions to decrease collisions and increase channel goodput~\cite{pcarma}, scheduling for reliable and efficient data collection~\cite{bhatia2020control}, and scaling~\cite{8516298}.
The validity and performance of the proposed improvements and solutions has been extensively tested in lab-like testbeds and simulations which have little in common with real-world environments and deployments. This is due to the absence of publicly available traces from the target environments.
The IoT research community has recognised the relevance of quantitative evidence from real-world environments and deployments and has made available a few LoRaWAN datasets~\cite{fingerprinting,10.1145/3359427.3361912,blenn2017lorawan}. 
The fingerprinting dataset~\cite{fingerprinting} contains over $120,000$~traces collected over a three month period. 
Each trace stores the location of the device, the Received signal Strength Indicator sampled by the gateways upon receiving a packet, the spreading factor used, and the time of the received packet. 
The fingerprinting dataset provides no gateway capacity, load, or deployment environment information. The LoRa underground link dataset~\cite{10.1145/3359427.3361912} is collected in an agricultural environment. It contains data from a low-density LoRaWAN deployment with only five transmitters and two receiver base stations. 

In this paper we present \loed, the LoRaWAN at the Edge Dataset, which consists of traces gathered in central London in a mix of dense urban and park environments. 
Overall, $11,000,000$~ packets (referred to as packets in this paper) were collected at nine gateways during 2019 and 2020.
We describe \loed in Section~\ref{sec:data}, and briefly discuss how the dataset can be exploited in Section~\ref{sec:usecases}. 
\section{Dataset}
\label{sec:data}

\fakeparagraph{Setup} \loed was acquired from nine LoRaWAN gateways in central London. The gateway locations were representative of typical dense urban and park environments and cover different deployment conditions as shown in Table~\ref{tab:gateways}.

Five outdoor gateways were deployed on the roof tops of large buildings, with a clear line-of-sight (LoS) to devices. Four indoor gateways were located near windows with limited LoS. One of the indoor gateway was placed on the ground floor of a college dormitory with no-LoS to any device. Each gateway forwarded received packets to a multiplexer which forwarded them to different Network Servers. Our server copied the packets and the gateway metadata to a time-series database. The gateway locations can be found in the dataset and table 1 contains information about the quantity of packets received at each gateway.
\begin{table*}
\small
\begin{tabular}{ p{0.12\textwidth} | p{0.9\columnwidth} || p{0.1\columnwidth} || p{0.2\columnwidth} | p{0.2\columnwidth} | p{0.2\columnwidth}}
\toprule
 Gateway ID & Description of location & Number of days & Total packets & max packets in a day & avg packets per day \\ 
 \midrule
00000f0c210281c4 & Dense outdoor area, on top of a building & 19 & 1326687 & 82781 & 69826 \\
00000f0c22433141 & Roof of a low building in a non-dense area & 36 & 144777 & 6579 & 4022 \\
00000f0c210721f2 & Top of a building in a very dense area and large open spaces & 56 & 5757575 & 121368 & 102814 \\
00000f0c224331c4 & Indoor in the ground floor a building, surrounded by buildings & 15 & 17029 & 1625 & 1135 \\
00800000a0001914 & Deployed inside a university building & 573 & 76706 & 2366 & 134 \\
00800000a0001793 & Deployed inside a university building & 552 & 186592 & 9596 & 338 \\
00800000a0001794 & Deployed inside a university building & 17 & 61080 & 4810 & 3593 \\
7276ff002e062804 & Deployed on top of a tall university building, with large open spaces & 131 & 1201916 & 15254 & 9175 \\
0000024b0b031c97 & Urban area, top of building, dense deployment & 131 & 2490639 & 25708 & 19013 \\
\bottomrule
\end{tabular}
\caption{Gateway data and location overview.}
\label{tab:gateways}
\vspace{-5mm}
\end{table*}
We collected LoRaWAN packets over a $2$-$4$~month period generated by smart city applications and research deployments.
The gateways received from $5,000$ up to $120,000$ packets per day. Overall, $11,263,001$ packets from $8,503$ unique device addresses with valid Cyclic Redundancy Check (CRC) and an uplink packet type were collected including packets with failed CRC. 

\fakeparagraph{The Data} We assumed that all collected packets were using the explicit header (EH) mode which enabled metadata extraction of: the device address, packet type, counter value, port number and other information like ADR. Upon packet reception at the gateway the timestamp, physical layer CRC status, frequency, spreading factor, bandwidth, coding rate, sampled RSSI and Signal to Noise (SNR) values are added to the packet information. Each trace consists of the following fields: 
\begin{table}[!htbp]
\begin{tabularx}{\textwidth}{ll}
\textbf{time} & Time at which the packet was received \\
\textbf{physical\_payload} & Raw payload received  \\
\textbf{gateway} & Gateway where the packet was received \\
\textbf{crc\_status} & Physical layer CRC \\
\textbf{frequency} & Radio frequency of the packet \\
\textbf{spreading\_factor} & LoRa spreading factor of the packet \\
\textbf{bandwidth} & Bandwidth of the received packet \\
\textbf{code\_rate} & LoRa coding rate of the packet\\
\textbf{rssi} & Sampled RSSI value at packet reception \\
\textbf{snr} & Sampled SNR value at packet reception \\
\textbf{device\_address} & Device address for the packet \\
\textbf{mtype} & Packet type \\
\textbf{fcnt} & Counter value of the packet \\
\textbf{fport} & Port of the packet \\
\end{tabularx}
\end{table}

The \loed dataset is publicly available at~\cite{dataset} and includes:
\begin{inparaenum}[$i)$]
\item all packets received at the nine gateways,
\item one pre-processed CSV data file for every day of the collection campaign. The files are saved in \textbf{dd\_mm\_yyyy.csv} format,
\item a set of scripts for processing and plotting the data and,
\item a preliminary analysis of the data. 
\end{inparaenum}

\fakeparagraph{Preliminary insights} 

\loed exposes insights into how LoRaWAN operates in real-world urban deployments and provides data such as:
\begin{inparaenum}[$a)$]
\item number of packets per day at a gateway,
\item total number of packets per node,
\item distribution of different packets types at gateways,
\item distribution of frequencies used at gateways,
\item distribution of spreading factors used at gateways,
\item distribution of RSSI values at a gateway,
\item distribution of SNR values at a gateway.
\end{inparaenum}

\begin{figure}[t!]
    \centering         
    \includegraphics[width=0.4\textwidth]{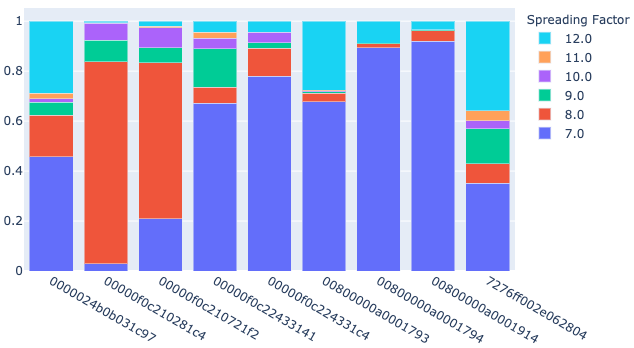}
    \vspace{-3mm}
    \caption{Distribution of spreading factor at each gateway.}
    \label{fig:sfdistribution}
    \vspace{-4mm}
\end{figure}

In Figure~\ref{fig:sfdistribution} we see the LoRa spreading factor usage at each gateway. Of note are the dominate use of spreading factors: 7, 8 (at two of the gateways) and 12. This may increase the probability of collisions as the number of devices using the same spreading factor rises. As a consequence, the performance of the applications running on the devices in the network may decrease and require further, lower-level, investigation. 

\section{Discussion}
\label{sec:usecases}

The \loed dataset can be used by the LoRaWAN community for many purposes. It can be used to 
characterise parameter usage and highlight long-term trends for LoRaWAN applications and devices
in an urban environment. \loed can provide test data to inform the design of scheduling 
algorithms and protocols, to ensure they are well-suited to the target applications and 
environments. Further, \loed can be plugged into capacity planning systems, which use different 
statistical and machine learning, to derive optimal parameters to improve network throughput, 
reduce interference or determine locations for new gateways to improve coverage.

\bibliographystyle{ACM-Reference-Format}
\bibliography{main}

\end{document}